\documentclass[reprint,amsmath,amssymb,aps]{revtex4-2}

\usepackage{graphicx}
\usepackage{dcolumn}
\usepackage{bm}
\usepackage{gensymb}

\begin{document}

\preprint{APS/123-QED}

\setlength{\abovedisplayskip}{1pt}
\setlength{\belowdisplayskip}{1pt}

\title{Low-Drift-Rate External Cavity Diode Laser}

\author{$^1$Eddie H. Chang$^\dagger$}
\thanks{These authors contributed equally.}
\author{$^1$Jared Rivera}
\thanks{These authors contributed equally.}
\author{$^1$Brian Bostwick}
\author{$^1$Christian Schneider}
\author{$^1$Peter Yu}
\author{$^{1,2,3}$Eric R. Hudson}
\affiliation{%
 $^1$Department of Physics and Astronomy, University of California, Los Angeles, California 90095, USA\\
 $^2$Center for Quantum Science and Engineering, University of California, Los Angeles, California 90095, USA\\
 $^3$Challenge Institute for Quantum Computation, University of California – Los Angeles, Los Angeles, California 90095, USA\\
 $^\dagger$Corresponding author: echang@physics.ucla.edu
}%

\collaboration{HUNTER Collaboration}

\date{\today}

\begin{abstract}
We present the design, construction, and simulation of a simple, low-cost external cavity diode laser with a measured free-running frequency drift rate of 1.4(1)~MHz/h at 852 nm. This performance is achieved via a compact, nearly monolithic aluminum structure to minimize temperature gradients across the laser cavity. We present thermal finite element method simulations which quantify the effects of temperature gradients, and suggest that the drift rate is likely limited by laser-diode aging.
\end{abstract}

\maketitle

The tunability and linewidth of external cavity diode lasers (ECDLs)~\cite{arnold_simple_1998}, combined with their relative ease and low cost of implementation, has made them a leading laser technology in  applications ranging from quantum information science~\cite{akerman2015universal} to bioscience~\cite{wang2012grating}.
For many of these applications, the laser frequency must be stabilized to $\lesssim1$~ppb over long periods of time.
This is typically accomplished by comparing the laser frequency to a stable frequency reference while using an active element, e.g. a piezoelectric transducer, to control the ECDL cavity length~\cite{hockel2009robust}. 
In these cases, a critical parameter is the free-running ECDL frequency drift rate, $\dot{\nu}$, which in combination with the mode-hop free tuning range of the ECDL sets how long the laser can remain locked to the frequency reference without user intervention.

Typical ECDL drift rates can be as large as several GHz/h~\cite{kobtsev_long-term_2008, koch2001frequency}, and primarily result from thermal cavity expansion induced by changes in environmental conditions~\cite{wieman_using_1991}.
To combat this, ECDLs often employ active temperature stabilization~\cite{clifford2001realization,zhao2004frequency} and/or athermal designs~\cite{talvitie1996technical, matthews1985packaged}.
In these cases, typical ECDL drift rates are a few tens of MHz/h.
We hypothesized that remaining drift is primarily due to temporal temperature gradients across the laser resulting in differential thermal expansion that cannot be corrected by active stabilization or an athermal design. 
This suggests the possibility of designing an ECDL to minimize temporal temperature \emph{gradients} across the laser cavity. 


Here, we present a home-built Littrow-style ECDL with improved drift over standard homebuilt designs. The laser is made out of a compact aluminum structure. 
While aluminum has a higher coefficient of thermal expansion than other materials used in laser construction, the ability of modern temperature controllers to stabilize to $< 1$~mK means that thermal-expansion-induced frequency drifts can be held to $\lesssim 1$ ppb.

Thus, the high thermal conductivity of aluminum can be utilized to minimize temperature gradients without degraded performance due to common-mode thermal expansion.
Similarly, a compact, nearly monolithic design, which includes a custom flexure mount, is employed to minimize temperature gradients. 
The compact design also results in improved acoustic immunity, higher mechanical resonance frequencies, and greater mode stability thanks to the larger external cavity mode spacing. With this design, we report a long-term drift rate of 1.4(1) MHz/h, which would allow a typical ECDL to operate free of mode hopping for roughly a year.

\begin{figure}[t!]
\includegraphics[scale=0.40]{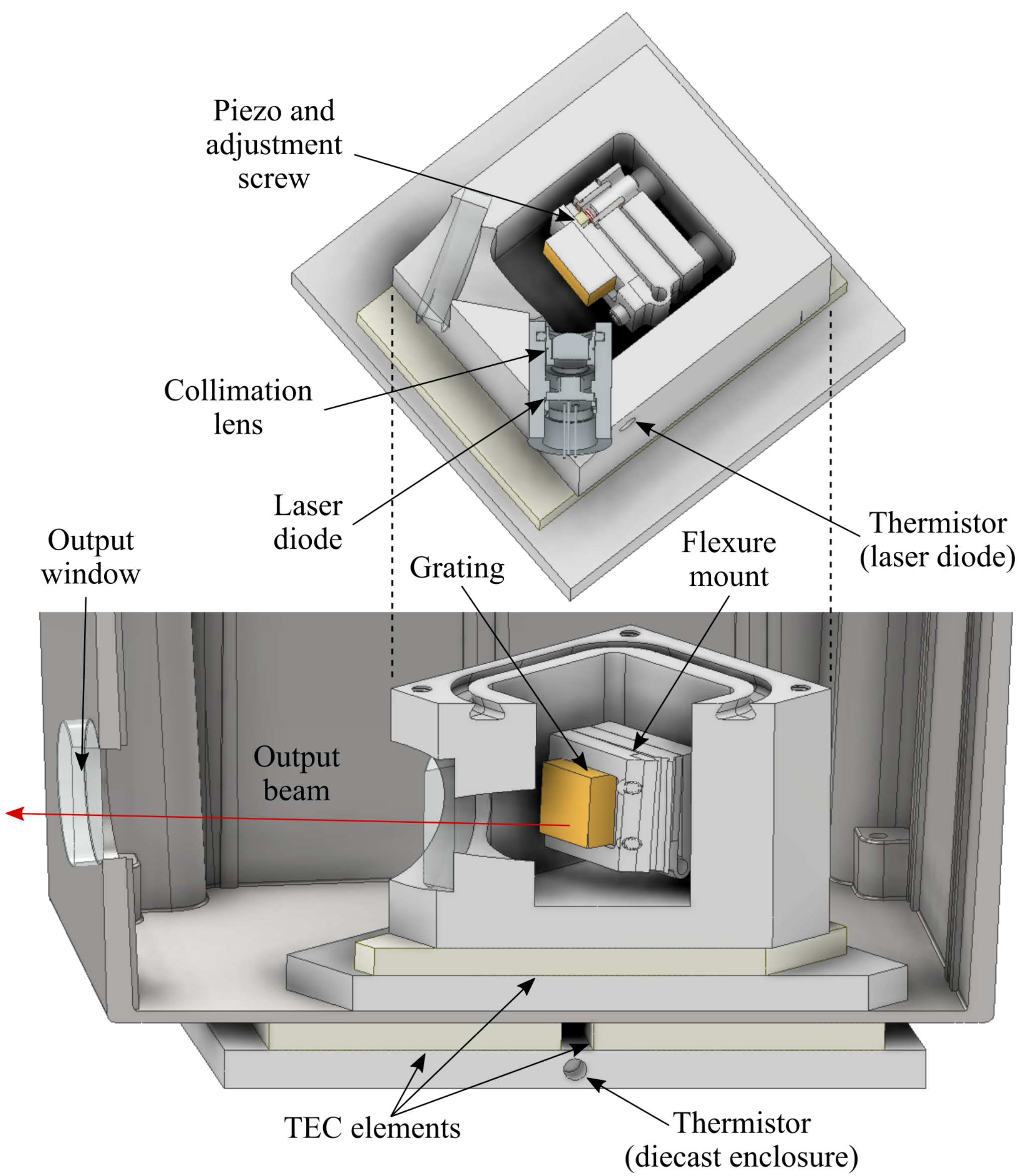}
\caption{\label{fig:XsecCAD} Cross-sectioned views of the present ECDL. }
\end{figure}

In what follows, we present the design and fabrication of the laser, followed by measurements of its performance. We conclude with the results of thermal finite element method (FEM) simulations to better understand the performance regarding temperature gradients.

Cross-sectioned views of the laser are shown in Fig.~\ref{fig:XsecCAD}.
The laser diode (Thorlabs, L852P150) is housed in a collimation tube (Thorlabs, LT230P-B) and collimated by an aspheric lens (Thorlabs, C330TMD-B, f = 3.1 mm).

The aluminum ECDL body holds the collimation tube at the Littrow angle ($\approx50^\circ$) with respect to the diffraction grating (Thorlabs GH13-18V), and the external cavity length is approximately 2.5~cm.
The output of the diode travels through a window in the ECDL and then through a window in a smooth diecast box, which houses the ECDL assembly. 
The beam is then passed through an optical isolator and a beam splitter, with one beam sent to a fiber-coupled wavemeter (HighFinesse WS-U, 10 MHz accuracy) and the other sent to an optical cavity for a side-of-fringe frequency noise measurement. 
The laser diode is powered by a custom-built current source with a current noise $< 150$~pA/$\sqrt{\textrm{Hz}}$ and output stability $<$~10 ppm/K.
The system is mounted on an optical table. 

The diffraction grating is epoxied onto a custom, back-adjust flexure mount which allows $\pm 4^\circ$ of horizontal and vertical adjustment. 
The mount is a three-piece assembly of high-strength 7075 aluminum. Coarse angular adjustment is provided by two 150 turns-per-inch screws inserted into precision bushings, which are press-fit into the mount. 
For fine angular adjustment, a piezoelectric transducer (PZT) rests at the end of the horizontal angle adjuster.
The PZT wires are fed out of airtight holes in the enclosure lid.

Both the ECDL assembly and the diecast box surface beneath are temperature controlled using 40~x~40~mm thermoelectric coolers (TEC), and a custom two-stage temperature control system that achieves $<$~0.2 mK/K stability (see Fig.~\ref{fig:XsecCAD}). 
Thermistors are epoxied into the ECDL enclosure and bottom-most baseplate to monitor the temperature of the ECDL assembly and beneath the diecast box, respectively. Their locations are indicated in Fig.~\ref{fig:XsecCAD}.

The ECDL enclosure thermistor wires are epoxied to the aluminum baseplate beneath to mitigate heat transfer from the outside. This was found to be necessary during tests of the temperature control system to achieve $<$~0.2 mK/K stability.
Thermally conductive epoxy was used to assemble the ECDL, as well as to fill in air gaps for improved thermal conductivity.
The ECDL assembly, diecast box, and electrical connectors feature rubber gasket seals which isolate the laser from air currents and changes in ambient pressure and humidity~\cite{park_compact_2003}.

\begin{figure}[!t]
\includegraphics[scale=0.65]{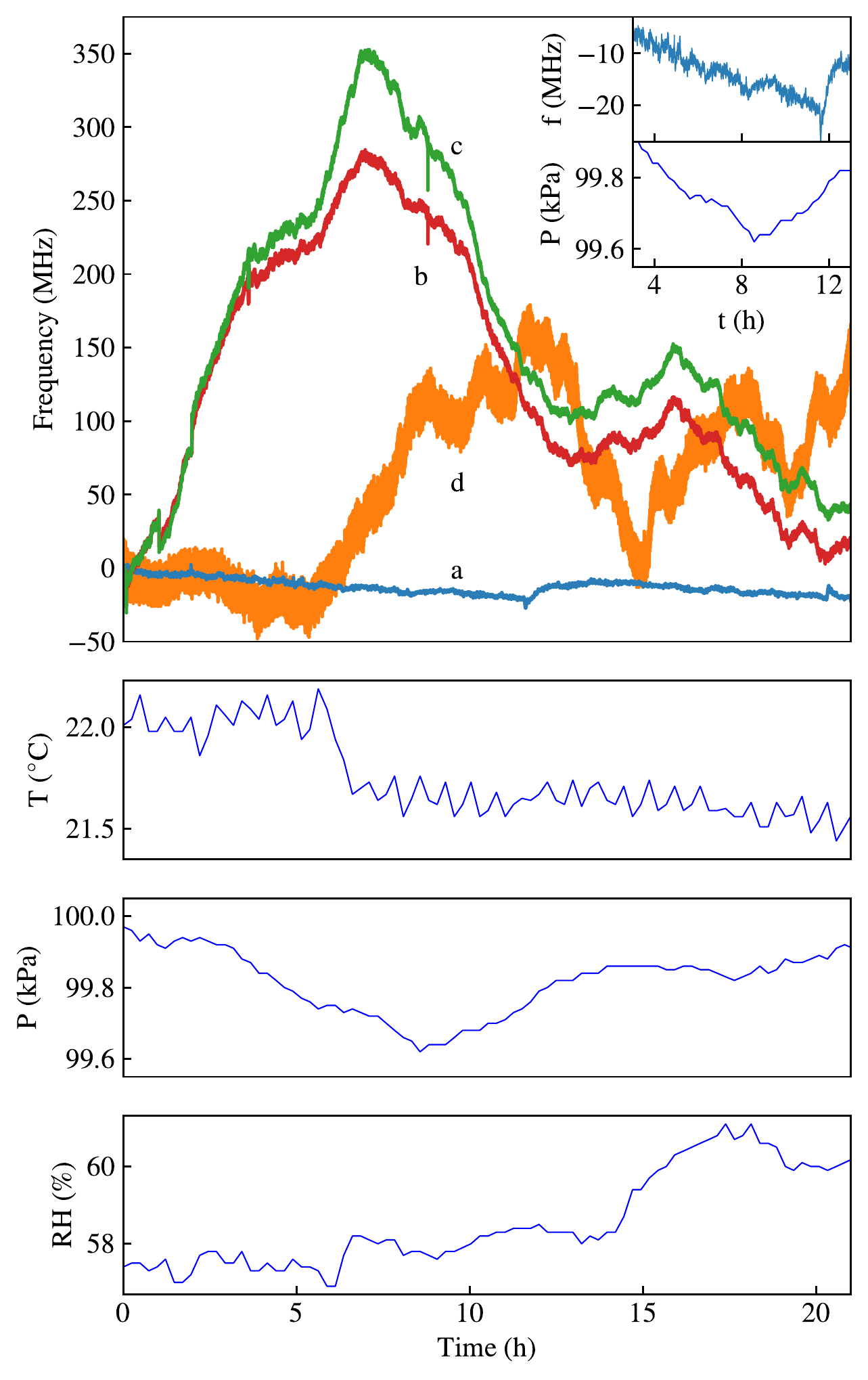}
\caption{\label{fig:Drift}Frequency drifts of the present ECDL (a), of commercial 650 nm (b) and 493 nm (c) ECDLs (Toptica, DL Pro), and of another home-built Littrow-style ECDL (d) monitored over 21 h. Beneath are the ambient temperature, pressure, and humidity during the run for (a, b, c). The inset juxtaposes the present ECDL frequency drift with a turning point in the ambient pressure at t = 8 h.}
\end{figure}

To measure the drift, the ECDL center frequency was recorded over a period of 22 hours using the wavemeter as shown in Fig.~\ref{fig:Drift}. The laser diode current was held at 98 mA, and the ECDL and diecast enclosures were held at 23.3\degree C and 24.3\degree C respectively.
Fig.~\ref{fig:Drift} also shows the laboratory temperature, pressure, and humidity during this time. 
To benchmark this performance, we also simultaneously measured the drift rates of two commercially available ECDLs (Toptica, DL Pro) (2b and 2c).
For further comparison another home-built ECDL (2d), a well-known Littrow-style design~\cite{hawthorn2001littrow}, was monitored over a similar duration, but on a different day.

The frequency changes of the benchmark lasers (2b, 2c, 2d) appear to be predominantly from pressure-induced fluctuations of a few hundred MHz.
The data for the present ECDL (2a) shows an overall frequency drift of -0.5(9) MHz/h over 21 h, where the 0.9 MHz/h error was obtained by including the 10 MHz absolute accuracy of the wavemeter. In an additional drift measurement free of wavemeter error, a beat-note center frequency with another ECDL locked to the cesium D${}_2$ transition was monitored. 
The beat note measurement resulted in a drift rate of 1.4(1) MHz/h over a period of 42 h, giving a fractional laser frequency drift of $\dot{\nu} = 4.0(3) \times 10^{-9}$/h. 

Different sources of drift were considered to understand the present design's free-running drift rate.
The laser tuning coefficients with respect to injection current and piezo tuning voltage are -375 MHz/mA and -30 MHz/V, respectively. We have measured our controllers' output stabilities with respect to ambient temperature to be 6.7 ppm/K and 0.2 mK/K. Assuming a typical daily temperature fluctuation of $1^\circ$C, each controller contributes $<$~0.1 MHz/h drift for the presented design, and cannot explain the observed drift.


Regarding ambient pressure and humidity fluctuations, first the Edl\'{e}n equation was used to estimate the change in refractive index with respect to pressure and relative humidity. For our typical weather parameters and wavelength of 852 nm, we find $\partial n_{air}/{\partial P} = 2.64 \times 10^{-9}~\text{Pa}^{-1}$ and $\partial n_{air}/\partial (RH\%) = -9.5 \times 10^{-9}~\text{(RH \%)}^{-1}$. Using the condition that the laser wavelength must change to absorb the change in optical length \cite{talvitie1996technical}, we find $\partial \nu / \partial \text{P}$ = -80 MHz/hPa and $\partial \nu / \partial \text{(RH\%)}$ = 3 MHz/(RH\%) for our laser - however, these figures hold only if the laser is fully exposed to the lab environment.

The actual isolation of the laser can be estimated using the turning point in ambient pressure at t = 8 h (Fig.~\ref{fig:Drift}, inset). The pressure rate change after passing the turning point is 1 hPa/h. The drift rate before the turning point is -2 MHz/h, and we suggest that a 1 MHz/h change after the turning point would be detectable. Such a change is not seen, implying the pressure in the laser enclosure changes by a factor $>$ 50 less. The laser is likely isolated even better against humidity fluctuations. Combining these points with the observed weather fluctuations in Fig.~\ref{fig:Drift}, we estimate the drift rates due to pressure and humidity fluctuations to be $<$ 0.4 MHz/h and $<$ 0.02 MHz/h, respectively.
Thus, the observed drift also cannot be explained by changes in ambient temperature or humidity.

\begin{figure}[h]
\includegraphics[width = .48\textwidth]{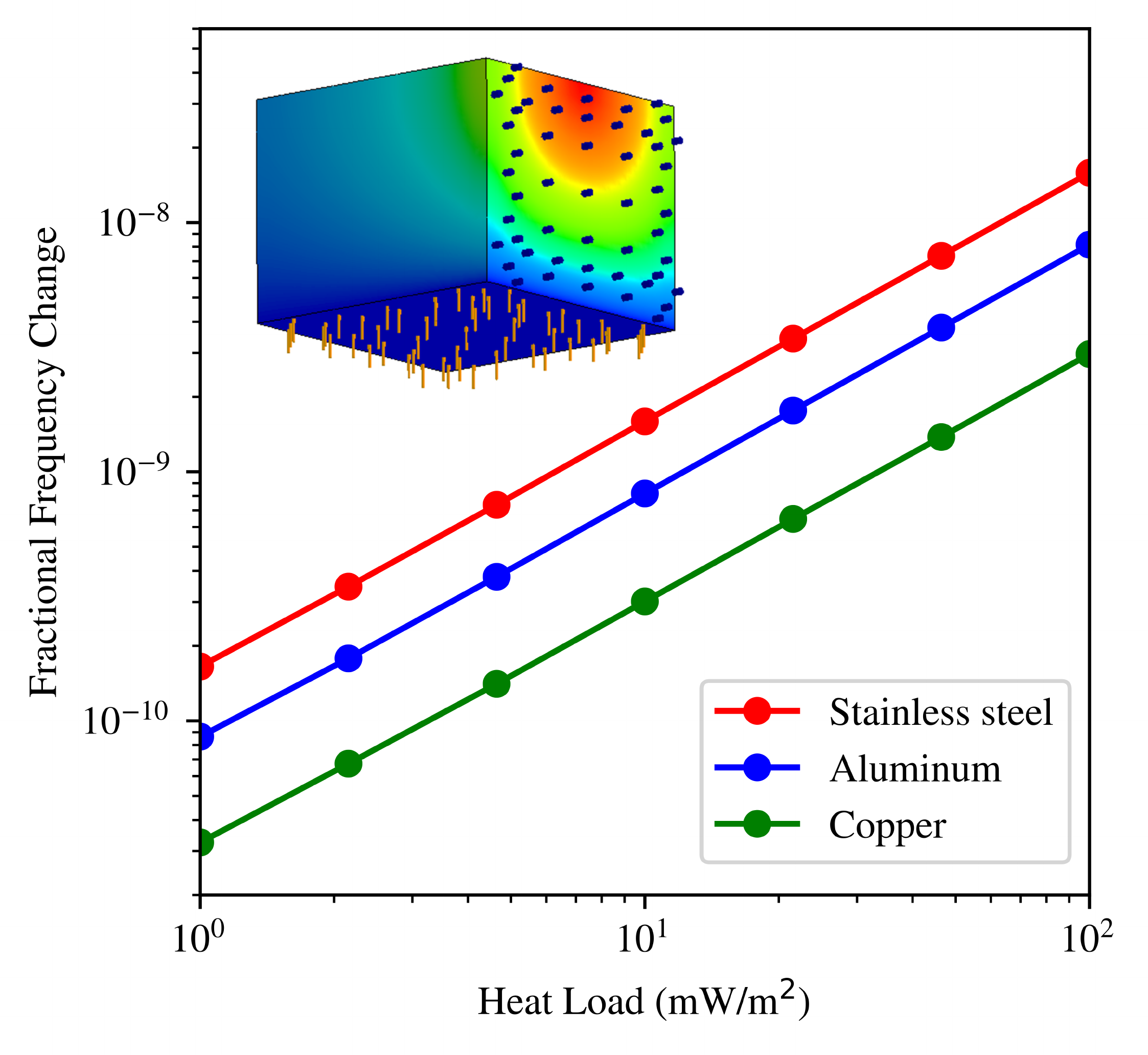}
\caption{\label{fig:Sim}Strain-related frequency change induced by a heat load on the ECDL enclosure. The cavity length change was obtained from thermal FEM of the enclosure (pictured). The bottom face was held at a constant temperature to model the TEC, and the heat load was applied to the front-and-back face pair. Out of the three materials simulated (stainless steel, aluminum, copper), copper produced the lowest frequency changes due to the resulting thermal gradients.}
\end{figure}

To understand how temperature gradients affect the laser frequency stability, FEM simulations were performed using a commercial package~\cite{SolidworksThermal}. Care was taken to ensure correct material definitions, thermal connections, absence of problematic geometries (e.g. thin shells, irregular edges, etc.) and mesh convergence. To model the effect of temperature gradients, the bottom of the ECDL assembly, which is in contact with the TEC, was held at a fixed temperature while a variable heat load was applied to one side of the assembly. The resulting frequency change was inferred from the cavity length change, and is plotted in Fig.~\ref{fig:Sim} for the pair of opposing faces which resulted in the largest strains. 

An upper-bound estimate of the actual heat load on the enclosure is the following. The laser experiences unequal blackbody radiation from the surrounding diecast walls. We have measured typical temperature differences between opposite diecast walls to be $\lesssim$~100~mK. Assuming emissivities for oxidized and polished aluminum for the diecast and ECDL enclosures respectively, and calculating the corresponding view factor, gives the upper bound $\lesssim$~1.5 mW/m${}^2$ of unequal heating. Fig.~\ref{fig:Sim} shows that this corresponds to a fractional frequency change of approximately $1\times 10^{-10}$, an order of magnitude below the measured hourly drift. This suggests that the long-term drift is no longer limited by temperature gradients in the cavity.


In addition to aluminum, stainless steel and copper ECDL assemblies were also simulated. Fig.~\ref{fig:Sim} shows that thermal effects in stainless steel were worse than in aluminum, and that copper outperformed both stainless steel and aluminum. This can be understood in terms of the materials' expansion coefficients and thermal conductivities. Compared to aluminum, stainless steel has lower expansivity but significantly lower conductivity, and copper has both lower expansivity and higher conductivity. In particular, copper appears to be an even better construction material for ECDL designs which aim to minimize drift due to thermal gradients.

Notably, drift rates as high as 30 MHz/h in Fabry–P\'{e}rot diodes have been reported \cite{wieman_using_1991}. As we report a lower drift rate, aging cannot be ruled out as the remaining source of drift. Facet oxidation and defect migration into the active region are both understood to contribute to drift in Fabry–P\'{e}rot diodes \cite{fukuda1994degradation}. In addition, a list of aging mechanisms that could cause drift in diode lasers has been compiled previously \cite{wieman_using_1991}.

Table 1 summarizes the discussed sources of frequency drifts and their estimated individual drifts for this ECDL under typical laboratory conditions.
Given the magnitude of the other effects considered here, the long-term frequency stability for the present design is likely limited by diode aging. 

\begin{table}[h!]
\begin{center}
\begin{tabular}{ |c|c|c| } 
 \hline
  & Drift \text{(MHz/h)} \\
 \hline
 ECDL temperature control stability & 0.1 \\ 
 \hline
 Current source stability & 0.01 \\
 \hline
 Ambient pressure fluctuations & $<$~0.4 \\ 
 \hline
 Ambient humidity fluctuations & $<$~0.02 \\
 \hline
 External cavity temperature gradients & $<$~0.1 \\
 \hline
 Laser diode aging & $\lesssim$~3 \\
 \hline
 Measured drift & 1.4(1) \\
 \hline
\end{tabular}
\caption{Estimated frequency drifts of the present ECDL. The drift due to external cavity temperature gradients was obtained by taking the FEM simulation frequency change associated with the estimated upper-bound heat load on the ECDL (see text).}

\label{table:1}
\end{center}
\end{table}

\begin{figure}[h]
\includegraphics[scale=0.68]{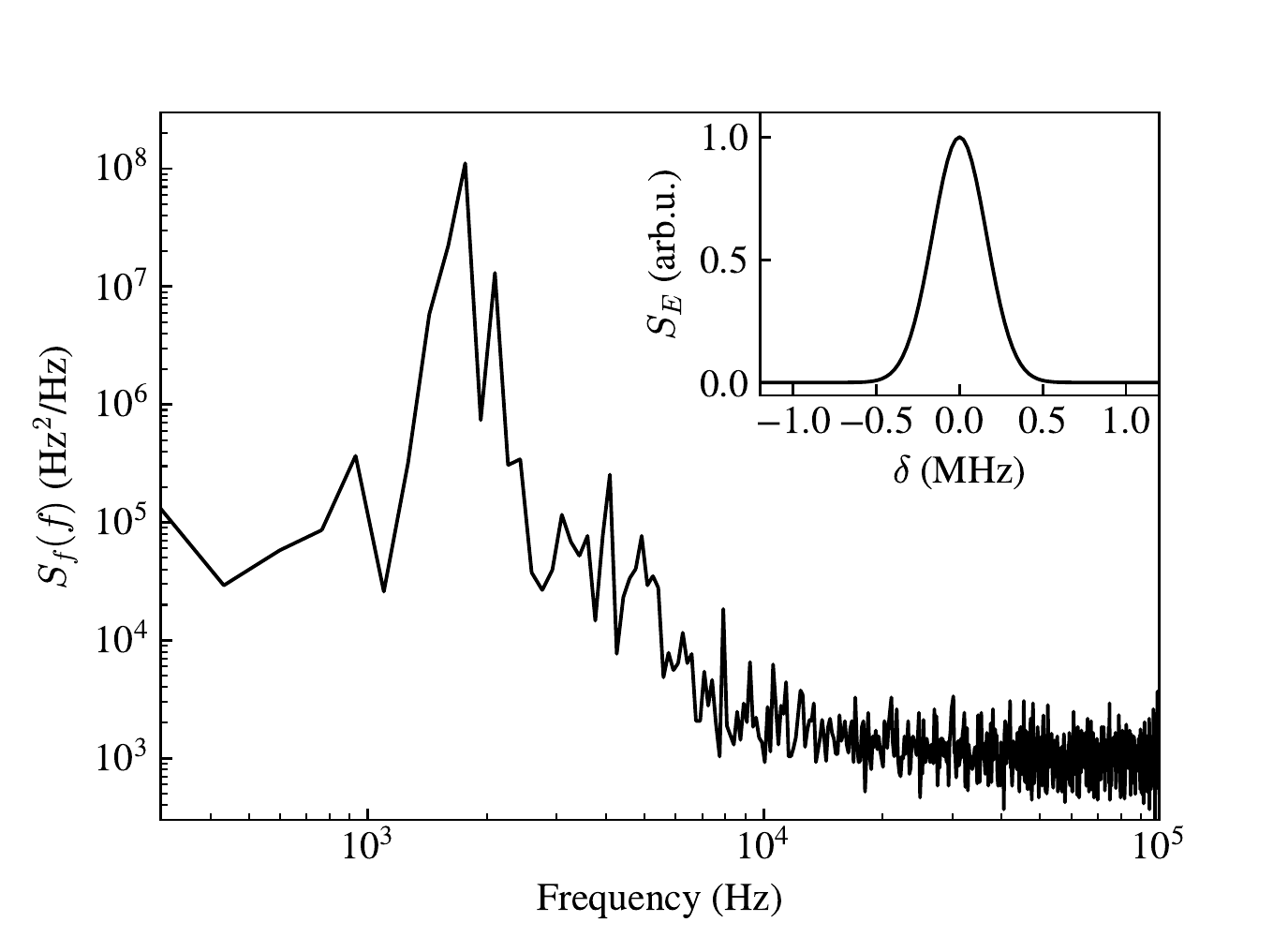}
\caption{\label{fig:LineWidth} Frequency noise power spectral density of the laser. The peaks near 2 kHz are acoustic resonances. The inset shows the calculated lineshape used to obtain the approximate linewidth of 400~kHz.}
\end{figure}

Finally, to further characterize the laser performance, its short-term linewidth was measured through a side-of-fringe method~\cite{coluccelli2015frequency}. The frequency noise power spectral density of the laser was measured and used to calculate the laser lineshape, as shown in Fig.~\ref{fig:LineWidth}. From the lineshape, the linewidth is found to be approximately 400~kHz.\\

In summary, we have described the design and construction of a simple and compact ECDL made from aluminum to minimize temperature gradients across the laser cavity.
The central frequency drift rate was measured to be 1.4(1) MHz/h, significantly smaller than other simple designs reported previously.
Thermal FEM simulations imply that the observed frequency instability is not caused by external cavity temperature gradients.
This points to diode aging effects as the likely limit on long-term frequency stability of this design. 

The approximate linewidth of 400~kHz makes this design useful for laser cooling and trapping applications.

\begin{acknowledgments}
This work was funded by the W. M. Keck Foundation and the Army Research Office.
We thank Justin Christensen, Matthew Boguslawski, and Hao Wu for their assistance with frequency drift measurements.
\end{acknowledgments}

\section*{Data Availability}
The data that support the findings of this study are available from the corresponding author upon reasonable request.

\bibliography{low_drift_ECDL_final_bib}

\end{document}